\documentstyle[aps,manuscript,floats]{revtex}

\input{epsf}

\begin{document}

\draft
\tighten

\title{
Almost-Hermitian Random Matrices:\\Eigenvalue Density
in the Complex Plane
      }
\author{
Yan V. Fyodorov$\S$\thanks{
On leave from Petersburg Nuclear Physics Institute, Gatchina
188350, Russia},
Boris A. Khoruzhenko$\P$\thanks{
On leave from B.I. Verkin Institute for Low Temperature Physics,
Kharkov, 310164, Ukraine.
       },
and Hans-J\"{u}rgen Sommers$\S$
       }
\address{
$\S$Fachbereich Physik, Universit\"at-GH Essen,
D-45117 Essen, Germany
        }
\address{
$\P$ School of Mathematical Sciences, Queen Mary \& Westfield
College, \\ University of London, London E1 4NS, U.K.
        }

\date{ June 24,1996}

\maketitle
\begin{abstract}
We consider an ensemble of large non-Hermitian random matrices of
the form
$\hat{H}+i\hat{A}_s$, where $\hat{H}$ and $\hat{A}_s$
are Hermitian statistically independent random $N\times N$
matrices. We demonstrate the existence of a new nontrivial
regime of weak non-Hermiticity characterized by the condition that
 the average of $N\mbox{Tr}  \hat{A}_s^2$ is of the same order
as that of $\mbox{Tr} \hat{H}^2 $ when
$N\to \infty$.
We find explicitly the density of
complex eigenvalues for this regime in the limit of infinite matrix
dimension. The density determines the eigenvalue distribution in
the crossover regime between random Hermitian matrices whose real
eigenvalues are distributed according to the Wigner semi-circle law
and random complex matrices whose eigenvalues are distributed in
the complex plane according to the so-called ``elliptic law''.
\end{abstract}

\section*{ }

Recently there was a growth of interest in statistics of complex
eigenvalues of large random matrices, both in physical and
mathematical literature
(see Refs.\ 1 - 20).

Non-Hermitian random Hamiltonians appear
naturally when one deals with quantum scattering problems in open
chaotic systems\cite{Sok1,Sok2,Haake,Lehm2,Dittes,FS}, studies
a motion of flux lines in superconductors with columnar
defects \cite{Nels} or is interested in chiral symmetry breaking in
quantum chromodynamics \cite{QCD}. Complex random matrices appear
also in
studies of dissipative quantum maps \cite{Grobe,Haake_book}.
On the other hand, closely related ensembles of real asymmetric
random matrices
 enjoy applications in neural network
dynamics \cite{Somp,Doyon} and in the problem of the localization
transition of random heteropolymer chains \cite{Nechaev}.

The actual progress in understanding the properties of random
matrices with complex eigenvalues is rather limited, however.
Most of the known results refer to the case of
{\it strong} non-Hermiticity  or asymmetry.
Namely, they deal with those
types of matrices for which the real
and  imaginary  parts of complex eigenvalues $Z_j=X_j+iY_j$
are typically of the same order when the matrix dimension tends
to infinity.

As the most well-known example, we mention
an ensemble of complex $N \times N$ matrices $\hat{J}$  with the
density of joint probability distribution of matrix entries of the
form
\begin{eqnarray}
{\cal P}(\hat{J})
=C_N^{-1} \exp \Big[-\frac{N}{J_0^2(1-\tau^2)} \ \mbox{Tr}
(\hat{J}\hat{J}^\dagger -\tau \ \mbox{Re}\  \hat{J}^2) \Big],
\label{P(J)}
\end{eqnarray}
where $\tau $ is real, $-1 \le \tau \le 1$,  and $C_N=[\pi^2
J_0^4(1-\tau^2)/N^2]^{N^2/2}$. In the large-$N$ limit the
eigenvalues of $\hat{J}$ are uniformly distributed in the ellipse $
X^2/(1+\tau )^2 + Y^2/(1-\tau )^2 \le J_0^2$ \cite{Gir2,Som,Lehm1}.

Denote by $J_{kl}$ a matrix element of $\hat{J}$. Eq.~(\ref{P(J)})
implies that: (i) complex variables $J_{kl}$  are Gaussian with mean
zero and variance $\langle|J_{kl} |^2  \rangle =J_0^2/N$, (ii) the
real and imaginary parts of $\hat{J}$ are statistically
independent, and (iii) among all the random variables $J_{kl}$, $1
\le k,l \le N$, only $J_{kl}$ and $J_{lk}$ are pairwise correlated
with the magnitude of correlations determined by the covariance
$\langle J_{kl}J^*_{lk} \rangle = \tau J_0^2 /N$. The ensemble
defined by Eq.~(\ref{P(J)}) interpolates between the well-known
Gaussian ensemble of
Hermitian  matrices (GUE) \cite{Mehta} and Ginibre's ensemble of
complex matrices \cite{Gin}. The degree of non-Hermiticity is
controlled by the absolute value of $\tau $. If $\tau =1$ or $\tau
=-1$ we have an ensemble of Hermitian or skew-Hermitian  matrices
and the above mentioned
 ellipse degenerates into an interval of real or purely
imaginary eigenvalues.
On the other hand, if $\tau =0$ then all $J_{kl}$ are mutually
independent
and we have a maximum of asymmetry. In that case the eigenvalues are
uniformly distributed in the unit circle \cite{Gin,Mehta,Gir1}.

The above mentioned ensemble of non-Hermitian random matrices can be
represented in another form. Each matrix $\hat{J}$ can be decomposed
into a sum of its Hermitian and skew-Hermitian parts:
$
\hat{J}=\hat{H}+i\hat{A}_s,
$
where  $\hat{H}=(\hat{J}+\hat{J}^\dagger )/2 $ and $\hat{A}_s
=(\hat{J}-\hat{J}^\dagger )/2i$. By Eq.~(\ref{P(J)}), the Hermitian
matrices $\hat{H}$ and $\hat{A}_s$ are statistically  independent
and follow
the   Gaussian distributions  ${\cal P} (\hat{H}) \propto \exp
\{-2N/[(1+\tau )J_0^2] \ \mbox{Tr}
\hat{H}^2 \} $ and ${\cal P} (\hat{A}_s) \propto \exp \{-2N/[(1-\tau
)J_0^2] \ \mbox{Tr}
\hat{A}_s^2 \}$, respectively. In this representation interpolation
mentioned above becomes even clearer. For instance, when $\tau $
approaches 1 variations of $\hat{A}$ around its mean value vanish
and $\hat{J}=\hat{H}$ in the limit of $\tau =1$.

For our purpose it is convenient to introduce new variables,
$J^2=J_0^2(1+\tau)/4$ and  $v^2=(1-\tau)/(1+\tau)$, and
rewrite the ensemble defined by Eq.~(\ref{P(J)}) as
\begin{eqnarray}\label{1}
\hat{J}=\hat{H}+iv\hat{A},
\end{eqnarray}
where now each $\hat{H}$ and $\hat{A}$
is taken from identical and independent Gaussian ensembles of
Hermitian matrices defined by the probability densities:
\begin{eqnarray}\label{2}
{\cal P}(\hat{H})\propto
\exp{\Big(-\frac{N}{2J^2}\mbox{Tr}\hat{H}^2 \Big)} \;\; \mbox{and} \; \;
{\cal P}(\hat{A})\propto
\exp{\Big(-\frac{N}{2J^2}\mbox{Tr}\hat{A}^2\Big)}.
\end{eqnarray}

It is natural to use the parameter $v^2$ as a measure of
non-Hermiticity (asymmetry) in our ensemble. As we mentioned,
in the case of strong non-Hermiticity,  i.~e.
when $v^2=O(1)$ as $N\to \infty$, the eigenvalues of $\hat{J}$
are uniformly distributed in an ellipse.

In the present paper we  demonstrate the existence of
a new non-trivial regime of weak non-Hermiticity
determined by the condition $v^2N=O(1)$ as $N\to \infty$. In other words,
we scale the parameter $v$ with the matrix dimension $N$
as $v=\alpha /\sqrt N$, $\alpha \ge 0 $.
In that regime of weak non-Hermiticity the
``elliptic law'' is not valid any longer and should be replaced by
another distribution. The main goal of our publication is to
derive the explicit form of this distribution.

In order to get better understanding of the chosen scaling
it is instructive to consider first the limiting case
of extremely weak non-Hermiticity. In this case  one
expects that the
influence of adding the non-Hermitian matrices $iv\hat{A}$ to the
Hermitian
matrices $\hat{H}$ in Eq.~(\ref{1}) can be treated by
regular perturbation theory.

When $v=0$ the eigenvalues $Z_j\equiv X_j$ of
$\hat{J}\equiv\hat{H}$ are real
and in the large-$N$ limit they are distributed according to the
Wigner semi-circle law with the density  $\rho_{sc}(X)=(2\pi J)^{-1}
\sqrt{4-(X/J)^2 }$.  The mean
separation $\Delta $ between adjacent eigenvalues can be estimated as
$\Delta \propto J/N$.
When a small non-Hermitian part $iv\hat{A}$ is
taken into account, the eigenvalues $Z_j$ of $\hat{J}$ move into
the complex plane
and their shift can be estimated by first order perturbation
theory. Namely, $\mbox{Re}\ Z_j=X_j$ and $\mbox{Im}\ Z_j=v \langle
\phi_j \!
\mid\hat{A}\mid \!\phi_j \rangle$, where $X_j$ and $\mid \! \phi_j
\rangle$   are the ordered  eigenvalues of $\hat{H}$ and their
normalized eigenvectors, respectively. Let us estimate the magnitude
of this shift. Introducing the notation
$\hat{P}_j$ for the projection on $\mid \! \phi_j \rangle$, one can
write
 $\mbox{Im}\ Z_j=v\mbox{Tr} \hat{A}\hat{P}_j $, so the variance of
 $Y_j\equiv \mbox{Im}Z_j$ is
\begin{eqnarray}\label{var1}
\langle \ Y^2_j  \ \rangle =
{\langle {\langle \ Y^2_j \ \rangle}_{\! A} \rangle}_{\! H}= v^2
{\langle {\langle \ \big|\mbox{Tr} \hat{A} \hat{P}_j \big|^2 \
\rangle}_{\! A} \rangle}_{\! H},
\end{eqnarray}
where the angle brackets ${\langle \cdot \rangle}_{H(A)} $ denote
the averaging over the random matrices $\hat{H} (\hat{A})$ according
to Eq.~(\ref{2}).
Since $\hat{H}$ is statistically independent of $\hat{A}$, so is
$\hat{P}_j$. As a result, one can perform the averaging over
$\hat{A}$ in Eq.~(\ref{var1}) explicitly:
\begin{eqnarray}\label{var2}
{\langle {\langle \ Y^2_j \ \rangle}_{\! A} \rangle}_{\!
H}=\frac{4J^2}{N}
{\langle  \ \mbox{Tr} \hat{P}_j^2 \ \rangle }_{\! H}.
\end{eqnarray}
Now noticing that $\hat{P}_j^2=\hat{P}_j$ and $\mbox{Tr} \hat{P}_j
= 1$ (as it must be for any projection), we obtain
 the deviation of $Y_j$, ${\langle \ Y^2_j \ \rangle}^{1/2}=
2vJ/N$. Since we are using perturbation theory, this deviation has
to be compared with the mean separation $\Delta $ of the
unperturbed
eigenvalues: $ {\langle \ Y^2_j \ \rangle}^{1/2}\! /\Delta \propto
v\sqrt N = \alpha $; so the case $\alpha  \ll 1$ corresponds to
well-defined perturbation theory.

The density $\rho (X,Y)$ of complex eigenvalues $Z_j$ can be
calculated easily in the perturbative regime $\alpha  \ll 1$.
Indeed,
$\rho(X,Y)=\langle\langle \ N^{-1} \sum_{j=1}^N
\delta(X-X_j)\delta(Y-Y_j) \ \rangle_{\! A}\rangle_{\! H}$.  Now
employing
 the Fourier representation for $\delta(Y)$, one finds that
\begin{eqnarray*}
\rho(X,Y)&=&\frac{1}{2\pi N} \sum_{j=1}^N \int_{-\infty}^{\infty}dk
e^{ikY}\Big\langle
\Big\langle \ \delta(X-X_j) \exp{\Big[\frac{-i\alpha k}{\sqrt N}\mbox{Tr}
\hat{A}\hat{P}_j\Big]} \ \Big\rangle_{\!\! A}\Big\rangle_{\!\! H}\\
 &=&
\frac{1}{2\pi N} \sum_{j=1}^N \int_{-\infty}^{\infty}dk
e^{ikY}\Big\langle \ \delta(X-X_j) \exp{\Big[
-\frac{(\alpha k
J/N)^2}{2}\mbox{Tr}\hat{P}^{\dagger}_j\hat{P}_j\Big]} \
\Big\rangle_{\!\! H}\\
 &=& \frac{1}{N} \sum_{j=1}^N
\big\langle \ \delta(X-X_j) \ \big\rangle_{\! H}
\frac{1}{2\pi }
\int_{-\infty}^{\infty}dk e^{ikY}
\exp{\Big[-\frac{1}{2}\Big(\frac{\alpha kJ}{N}\Big)^2  \Big]}
 \\
 &=& \rho_{sc}(X) \rho(Y),
\end{eqnarray*}
where
\begin{eqnarray}\label{3}
\rho(Y)=\frac{1}{\sqrt{2\pi}\tilde{a}}\exp
\left(-\frac{Y^2}{2\tilde{a}^2}\right)
\end{eqnarray}
and  $\tilde{a}=\alpha J/N$.
We see that for $\alpha \ll 1$ the density of eigenvalues in the
complex plane $X+iY$ is a simple product of the Wigner semicircular
density
$\rho_{sc}(X)=(2\pi J)^{-1} \sqrt{4-(X/J)^2}$
and the Gaussian $\rho(Y)$ from Eq.~(\ref{3}).
On the other hand, it is natural to
expect that when $\alpha\gg 1$ $\rho(X,Y)$ goes back to the
mentioned uniform distribution inside an ellipse.

In order to get access to the distribution of eigenvalues in the
complex plane $E=X+iY$ in the case of $\alpha=O(1)$ we use the fact
that the two-dimensional
density of these eigenvalues $\rho(X,Y)$ can be found if one knows
the ``potential''  \cite{Som}:
\begin{eqnarray}\label{4}
\Phi(X,Y,\kappa)=\frac{1}{2\pi
N}\left\langle\left\langle\ln{Det[(E-
\hat{J})(E-\hat{J})^{\dagger}+\kappa^2\hat{I}]}\right\rangle_{\! \!
A}\right\rangle_{\! H}
\end{eqnarray}
in view of the relation: $\rho(X,Y)=
\lim_{\kappa\to 0}\partial^2 \,\Phi(X,Y,\kappa)$, where
$\partial^2$ stands for the two-dimensional Laplacian
$\partial_X^2+\partial_Y^2$.
To determine the potential for the present case of almost-Hermitian
Gaussian random matrices we follow the method suggested earlier by
two of us \cite{FS}
and restore the potential from its derivative:
\begin{eqnarray}\label{5}
\frac{\partial^2 \Phi}{\partial \kappa^2}
 &=&
\frac{1}{2\pi N}
\frac{d}{d\kappa}
\lim_{\kappa_{b}\to\kappa}
\frac{\partial}{\partial \kappa}
\left\langle
\left\langle
\ln{Z(\kappa_b,\kappa)}
\right\rangle_{\! \! A}
\right\rangle_{\!\! H}
\\ \nonumber
Z(\kappa_b,\kappa)&=&
\frac{Det[(E-\hat{J})(E-\hat{J})^{\dagger}+\kappa^2\hat{I}]}
{Det[(E-\hat{J})(E-\hat{J})^{\dagger}+\kappa_{b}^2\hat{I}]} .
\end{eqnarray}
It is convenient for our purpose to write down the determinants in
the denominator and numerator of the generating function as:
\begin{equation}\label{denom}
Det[(E-\hat{J})(E-\hat{J})^{\dagger}+\kappa_{b}^2\hat{I}]=
Det\left[\hat{M_b}\left(\begin{array}{cc}\kappa_b \hat{I}& i(E-\hat{J})\\
i(E-\hat{J})^{\dagger}&\kappa_b
\hat{I}\end{array}\right)\hat{M}_b^{-1}\right]
\end{equation}
\begin{equation}\label{numerator}
Det[(E-\hat{J})(E-\hat{J})^{\dagger}+\kappa^2\hat{I}]=
Det\left[\hat{M}\left(\begin{array}{cc}
i(E-\hat{J})^{\dagger}&\kappa \hat{I}\\
\kappa\hat{I} &i(E-\hat{J})\end{array}\right)\hat{M}^{-1}\right]
\end{equation}
where we used two matrices:
$\hat{M}_b=\frac{1}{\sqrt{2}}\left(\begin{array}{cc}\hat{I}&-i\hat{I}\\
\hat{I}&i\hat{I}\end{array}\right);\quad
\hat{M}_b^{\dagger}=\hat{M}_b^{-1}$
and $\hat{M}=\frac{1}{\sqrt{2}}\left(\begin{array}{cc}\hat{I}&-\hat{I}\\
\hat{I}&\hat{I}\end{array}\right);\quad \hat{M}^{\dagger}=\hat{M}^{-1}$.

Then one can find the following convergent representation for the
generating
function $Z(\kappa_b,\kappa)$ in terms of a Gaussian integral over
both commuting and anticommuting (Grassmannian) variables:
\begin{eqnarray} \nonumber
(-1)^N Z(\kappa_b,\kappa)&=&\int[d\Psi]\exp\{-{\cal L}_0 (\Psi)-{\cal
L}_1 (\Psi)\}\\ \label{6}
{\cal L}_0 (\Psi)&=&
\kappa_{b}(\Psi^{\dagger}\hat{\Lambda}\hat{L}\Psi)+iX(\Psi^{\dagger}
\hat{L}\Psi)-Y(\Psi^{\dagger}\hat{\sigma}_{0}\Psi)+(\kappa-\kappa_b)
(\Psi^{\dagger}\hat{K}\Psi)
\\ \nonumber
{\cal L}_1
(\Psi)&=&-i\Psi^{\dagger}(H\otimes\hat{L})\Psi-
\frac{\alpha}{N^{1/2}}\Psi^{\dagger}
(\hat{A}\otimes\hat{\sigma}_{0})\Psi ,
\end{eqnarray}
where $\Psi^{\dagger}=(\vec{S}^{\dagger}_{1},\vec{S}^{\dagger}_2,
\chi^{\dagger}_1,\chi^{\dagger}_2);\quad
[d\Psi]=\prod_{p=1,2}{d\vec{S}_p d\vec{S}^{\dagger}_{p}
d\chi_p d\chi^{\dagger}_p}$, with $\vec{S}_p$ and $\chi_p$ being
$N-$component vectors of complex commuting and Grassmannian
variables, respectively.
The $4\times 4$ matrices $\hat{\Lambda},\hat{L},\hat{\sigma}_0$ and
$\hat{K}$
are (block)diagonal of the following structure:
$$
\begin{array}{r r r r r r}
\hat{\Lambda}&=&\mbox{diag}(1,-1,1,-1);&
\hat{L}&=&\mbox{diag}(1,-1,1,1); \\
\hat{K}&=&\mbox{diag}(0,0,1,-1);&\hat{\sigma_0}&=&
\mbox{diag}(i\Sigma_x,\Sigma_x);
\end{array}
$$
and
$
\Sigma_x=
\left(
\begin{array}{cc}
0&1\\1&0
\end{array}
\right)
$.

Due to the normalization condition $Z(\kappa,\kappa)=1$ it is enough for
our purpose to calculate the average $\left\langle\left\langle
Z(\kappa,\kappa_b)\right\rangle_{\! \! A}\right\rangle_{\! H}$.
This can be
 done easily for the Gaussian distributions
Eq.~(\ref{2}) and
leads to the terms quartic in the components of the supervector $\Psi$ in
the exponent of the integrand, Eq.~(\ref{6}). Further evaluation goes
along lines suggested by Efetov in the theory of disordered systems
\cite{Efe}.
A detailed introduction to the method as applied to random
Hermitian matrices can be found in the review\cite{my}. Here we
outline only the general strategy:
1) To decouple quartic terms in the exponent of the integrand of
the generating function by introducing a set of auxiliary
integrations (the so-called Hubbard-Stratonovich transformation, see
\cite{my}); 2) To integrate out
the $\Psi$-variables explicitly; and  3) Exploiting the limit $N\to
\infty$
to integrate out some ("massive") degrees of freedom in the
saddle-point approximation. After this is done the integral over
the remaining ("massless")
degrees of freedom can be represented in a form of the so-called
zero-dimensional graded nonlinear $\sigma-$model introduced into
physics by Efetov\cite{Efe}.

After this set of standard manipulations one arrives at the
following expression for the density $\rho_X(y)$ of the scaled
imaginary parts
$y_j=2\pi{\rho_{sc}(X)}Y_jN$ of those eigenvalues
$Z_j=X_j+iY_j$ whose real parts $X_j$ fall within a narrow window
around the point $X$ of the spectrum:
\begin{eqnarray}\label{7}
\rho_X(y)=\frac{1}{2}\frac{\partial^2}{\partial y^2}\int_{0}^{\infty}
du u\phi(y,u) ,
\end{eqnarray}
where
\begin{eqnarray}\label{8}
\phi(y,u)= i \displaystyle{\frac{\partial}{\partial u}}
\int d\mu(Q)\mbox{Str}(\hat{K}\hat{Q}) \,
\times\exp
\Big[-\frac{iu}{2}\mbox{Str}(\hat{Q}\hat{\Lambda})-\frac{iy}{2}
\mbox{Str}(\hat{Q}\hat{\sigma})-
\frac{a^2}{16}\mbox{Str}(\hat{Q}\hat{\sigma}\hat{Q}\hat{\sigma})  \Big],
\end{eqnarray}
where $a=2\pi{\rho_{sc}(X)}\alpha$,
$\hat{\sigma}=\mbox{diag}(\Sigma_x,\Sigma_x)$ and the
(graded) matrices
$\hat{Q}$ satisfying  $\hat{Q}^2=-1$ are taken from the graded
coset space $U(1,1/2)/U(1/1)\otimes U(1/,1)$,
whose explicit parameterization
can be found in \cite{Efe,my}. We also use the notation  Str  for
the graded trace.

Still, the evaluation of the integral over the graded coset space
in Eq.~(\ref{8}) and subsequent restoration of the density $\rho_X(y)$
is quite an elaborate task. We skip
intermediate steps and present the final expression for $\phi(y,u)$
which one obtains
after integrating out the Grassmannian variables:
\begin{eqnarray}\label{9}
\phi(y,u)&=&\langle I(z_1,z_2)\rangle_Z\equiv
\frac{1}{2\pi}\int_{-\infty}^{\infty}\int_{-\infty}^{\infty}dz_1dz_2
\exp{\left[-\frac{1}{2}(z_1^2+z_2^2)\right]}I(z_1,z_2)
\\ \nonumber
I(z_1,z_2)&=&y
\left[F_2\partial_yF_1-F_1\partial_yF_2\right]-\frac{a^2}{2}\left[
F_1F_2+\partial_yF_1\partial_yF_2
+\partial_uF_1\partial_uF_2-F_2\partial^2_yF_1
-F_1\partial^2_yF_2\right]
\end{eqnarray}
where
\begin{eqnarray}\label{10}
F_1(y,u,z_1,z_2)=\frac{e^{-\sqrt{u^2+y_b^2}} }{\sqrt{u^2+y_b^2}}; \quad
F_2(y,u,z_1,z_2)=\frac{ \sinh \sqrt{u^2+y_f^2} }{\sqrt{u^2+y_f^2}};
\end{eqnarray}
and $y_b=y-az_1/\sqrt 2;\,\,y_f=y-iaz_2/\sqrt 2$.

It is easy to check that both functions $F_1$ and $F_2$ satisfy the
relation:
\begin{eqnarray}\label{11}
\partial^2_yF_{1,2}=F_{1,2}-\frac{1}{u}\partial_u \left(u\partial_u
F_{1,2}\right) .
\end{eqnarray}

Now substitute Eq.~(\ref{9}) into Eq.~(\ref{7}) and notice  (by
multiple
exploitation of Eq.~(\ref{11}) ) that  $I(z_1,z_2)$ can be
written as a
full derivative over $u$. As a result, one obtains:
\begin{eqnarray}\label{12}
\rho_X(y)&=&
\Big\langle
[2+y\partial_y]
\left[
u(F_1
\partial_uF_2-F_2\partial_uF_1)
\right] \left\vert_{u\to 0} \right.
\Big\rangle_{\! Z} - \\ \nonumber
& &
\Big\langle
\frac{a^2}{2}\partial^2_y[F_1\partial_u(u\partial_uF_2)]
\left\vert_{u\to 0} \right.
-\frac{a^2}{2}\partial_y[u(F_2\partial^2_{uy}F_1-
\partial_yF_1\partial_uF_2)]
\left\vert_{u\to 0} \right.
\Big\rangle_{\! Z},
\end{eqnarray}
where $\langle ...\rangle_Z$ denotes the integration over $z_1,z_2$
with the Gaussian weight, see Eq.~(\ref{9}).
When performing the limiting procedure $u\to 0$ in Eq.~(\ref{12}).
one notices that all terms are proportional to a $\delta$-function:
$\delta(y_b)$,
or its derivatives. This fact allows one to perform the integration
over $z_1$ explicitly, and after some simple algebraic
manipulations one arrives at the following expression:
\begin{eqnarray}\label{13}
\rho_X(y)=\frac{1}{\sqrt{2 \pi} a}\exp \left( -\frac{y^2}{2a^2} \right)
\int_0^1  dt \cosh (ty) \exp{(-a^2t^2/2)},
\end{eqnarray}
where $a=2\pi \rho_{sc}(X) \alpha$ and $y=2\pi \rho_{sc}(X)YN$, $X$
and $Y$ being the real and imaginary parts of the complex
eigenvalues in the ensemble of random matrices $\hat{J}$ given by
Eqs.~(\ref{1})-(\ref{2}).

The distribution Eq.~(\ref{13}) constitutes the main result of the
present publication. It correctly reproduces all the anticipated
limiting cases.
When $a\gg 1$ one can effectively put the upper boundary of
integration in Eq.~(\ref{13}) to be infinity due to the Gaussian
cut-off of the integrand at $t\sim 1/a\ll 1$. This immediately
results in the uniform density $\rho_X(y)
=(2a^2)^{-1}$ inside the interval $| y|\le a^2$ and vanishing
density outside this interval. Recalling  the definition of the
variable $y$ and the parameter $a$, we
are back to the familiar ``elliptic law''. In the opposite limiting
case $a\ll1$ one obtains:
\begin{eqnarray*}
\rho_X(y)=\frac{1}{(2\pi)^{1/2}a}\exp \Big( -\frac{y^2}{2a^2} \Big)
\frac{\sinh{y}}{y}
\end{eqnarray*}
which matches the perturbative result Eq.~(\ref{3}) as long as $y\ll
1$. The fact that remote tails $y\gtrsim1$ are not captured
correctly by perturbative treatment can be understood easily: the
condition $y\sim 1$ means that the imaginary parts $Y_j$ of the
complex eigenvalues is of the same order as the mean  spacing
$\Delta=(\rho_{sc}(X)N)^{-1}$ between the real eigenvalues of the
unperturbed Hermitian matrix  $\hat{H}$. First order perturbation
theory is obviously insufficient to obtain
 those complex eigenvalues correctly.

It is worth mentioning that we have restricted ourselves to  Gaussian
distributions of random matrices $\hat{H}$ and  $\hat{A}$
mainly for the sake of simplicity of presentation of the
supersymmetry calculations.
More general distributions of matrices  $\hat{H}$ and  $\hat{A}$
can be considered and supersymmetry calculations can be performed
along the line of \cite{MF,FS1}. For instance one can consider
random matrices $\hat{H}$ and $\hat{A}$ whose entries are
independent Bernoulli variables taking $\pm b$ with equal
probability or are uniformly distributed in an interval.
 Actually, we have arguments that the universality
of the eigenvalue distribution in the crossover regime is extremely high
and is not restricted to the case when the corresponding Hermitian
matrices $\hat{H}$ have the semi-circle eigenvalue density.
For example, one can consider non-Gaussian distributions of
matrices $\hat{H}$ and $\hat{A}$ with the density  of the form
${\cal P} (\hat{H}) \propto \exp \{-N \mbox{Tr}
V(\hat{H}) \} $ (the so-called invariant ensembles, see
\cite{inv,BIZ,Pas1,BZ,DiFGZ,HW,PS}).
Then the formula Eq.~(\ref{13}) is still valid, provided the
semi-circular
density $\rho_{sc}$ in the definitions of $y$ and $a$ is replaced
by the actual
mean level density at given point of the spectrum\footnote{
 We assume that the matrices $\hat{H},\hat{A}$ normalized
in a way ensuring $\langle \mbox{Tr} \hat{H}^2\rangle=\langle
\mbox{Tr} \hat{A}^2\rangle=J^2\! N$.} .

\epsfxsize=16.5cm
\begin{figure}
\epsfbox{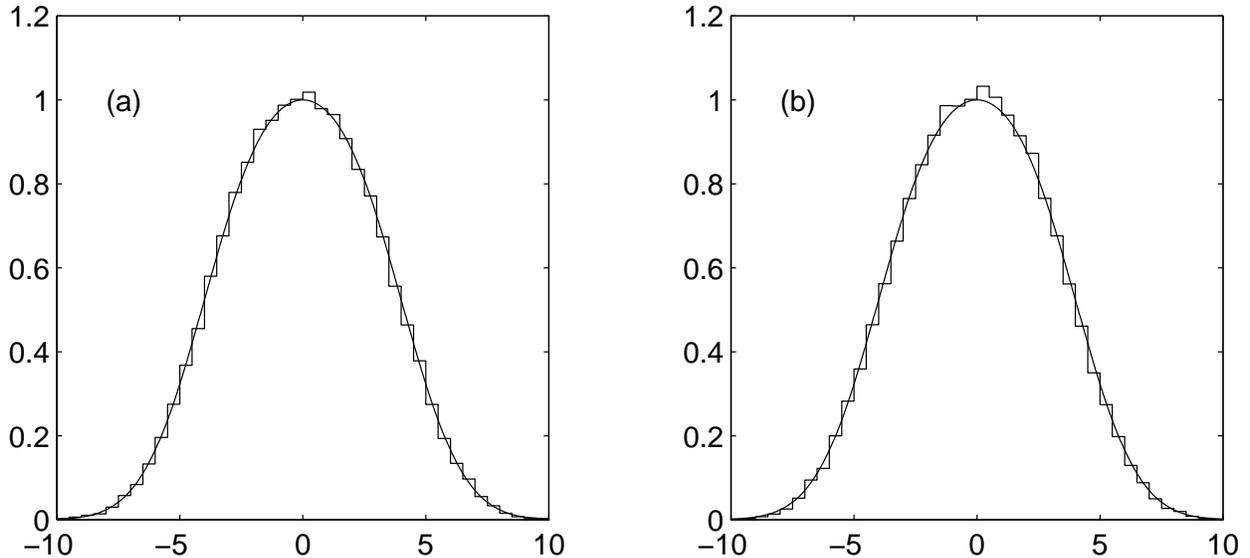}
\caption{Histograms of distributions of the scaled imaginary parts
of complex eigenvalues for a) Gaussian and b) Bernoulli distribution of
matrix elements, see the text.}
\end{figure}

The distribution Eq.~(\ref{13}) has been derived in the limit of
infinite
matrix
dimension. However, numerical computations show that Eq.~(\ref{13})
well approximates the distribution of complex eigenvalues for
ensembles of random matrices (\ref{2}) of a moderate dimension.
In Fig.~1 we present results of numerical diagonalization
of random matrices of dimension $N=30$. For each plot we generated
a set of
20000 random matrices $\hat{J}=\hat{H}+i\frac{1}{\sqrt N}\hat{A}$ sampling a
Gaussian distribution (random matrices  follow distributions of Eq.~(\ref{3})
with
$2J^2 = 1$) for plot (a) and a  Bernoulli distribution
(functionally independent matrix elements of $\hat{H}$ and $\hat{A}$
are statistically independent random variables taking $\pm
1/\sqrt{2N}$ with equal probability) for plot (b).
Then  matrices were diagonalized and their eigenvalues
$Z_j$ falling
into a small energy window, $|\mbox{Re}\ Z_j | \le 0.2$, were
selected. The imaginary parts  of the selected  eigenvalues were
counted and the corresponding histograms along with the infinite-$N$
theoretical density  $\rho_{X=0}(y)$ (solid curves) were plotted
against the scaled variable $y_j=2\pi{\rho_{sc}(0)}Y_jN$.
In both plots, the ordinate is scaled in units of $\max_y
\rho_{X=0}(y)$.

It is worth mentioning that our derivation based on the
supersymmetry formalism \cite{Efe,my} being in general quite
satisfactory should be considered as a heuristic one from the
point of view of rigorous mathematics. In a more detailed publication
\cite{prep} we shall give a rigorous mathematical derivation of our
main result, Eq.~(\ref{13}) for the case of Gaussian Matrices
$\hat{H},\hat{A}$
and also show the mentioned universality of the found distribution
for several
classes of more general random matrices.

At the moment we failed to derive the analogous
expression for
almost-symmetric (slightly asymmetric) real random matrices due to
unsurmountable technical problems. At the same time
there are reasons to suspect that for the latter case the distribution
might be different in its form from Eq.~(\ref{13}). It would be
of much interest to try to attack this problem by different methods,
e.g. starting from the known joint probability density for
eigenvalues  of asymmetric
random matrices\cite{Lehm1,E}.

In conclusion, we derived the explicit expression describing the
density of complex eigenvalues of almost-Hermitian random
matrices.
It describes the crossover regime from the Wigner semi-circle law
typical for random Hermitian matrices to the ``elliptic law''
typical for strongly non-Hermitian matrices.

The authors acknowledge financial support of Deutsche
Forschungsgemeinschaft under Grant No SFB237. Numerical computations
were performed at Institut f\"ur Mathematik,
Ruhr-Universit\"at-Bochum with the help of the MATLAB software package.


\end{document}